\documentclass{PoS}

\usepackage{psfrag,epsfig,graphicx,graphics,xcolor}
\usepackage{pstricks}
\usepackage{epsf}
\usepackage{rotating}
\title{QCD collinear factorization, its extensions and 
the partonic distributions}

\ShortTitle{QCD collinear factorization, its extensions and 
the partonic distributions}

\author{\speaker{Lech Szymanowski}%
    \thanks{I am thankful to the organizers of this conference for inviting me to deliver this talk and to B.~Pire and S.~Wallon for discussions on issues presented in my review.
This work is partly supported by the Polish Grant NCN
No DEC-2011/01/D/ST2/02069, the French-Polish collaboration agreement Polonium, the
P2IO consortium and the Joint Research Activity "Study of Strongly Interacting Matter"
(acronym HadronPhysics3, Grant Agreement n.283286) under the Seventh Framework Programme of the European Community.}
    \\
        National Center for Nuclear Research, Warsaw\\
        E-mail: \email{Lech.Szymanowski@fuw.edu.pl}}

\abstract{
I review the basics  of the collinear factorization theorem applied primarily to deep inelastic scattering (DIS) involving forward parton distributions (PDFs) and the extensions of this theorem for exclusive processes  probing non-forward parton distributions (GPDs), the generalized distribution amplitudes (GDAs) and the transition distribution amplitudes (TDAs). These QCD factorization theorem is an important tool in the description of hard processes in QCD. Whenever valid, it permits to represent the cross section or the scattering amplitude  for such a process  as a convolution in partonic momenta of a perturbatively calculable part (the coefficient function, CF) which involves the hard scale of the process with  non-perturbative (soft) distributions of active partons inside the hadrons involved in a process. The reliability of the perturbatively determined hard part  together with  high precision experimental data on relevant observables gives a hope for the possibility to uncover fine details of interpartonic interactions and  spatial distributions of partons inside hadrons.
I conclude with some remarks about  QCD factorization with transverse momentum dependent PDFs.
   
          }

\FullConference{Sixth International Conference on Quarks and Nuclear Physics\\
		 April 16-20, 2012\\
		 Ecole Polytechnique, Palaiseau,  France}

\begin{document}

%\section{...}

\begin{figure}[h!]
\begin{center}
\epsfxsize=0.7\textwidth
\epsffile{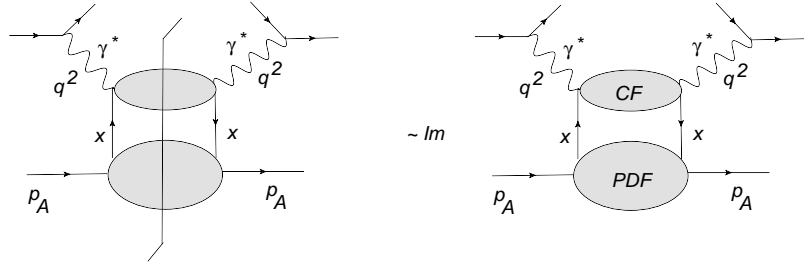}
\end{center} 
\caption{The parton model description of inclusive DIS (left figure) and its relation to the discontinuity of the forward scattering amplitude of an exclusive process (right figure).}
\label{figDISnew2amplhard}
\end{figure}

\begin{figure}[h!]
\begin{center}
\epsfxsize=0.7\textwidth
\epsffile{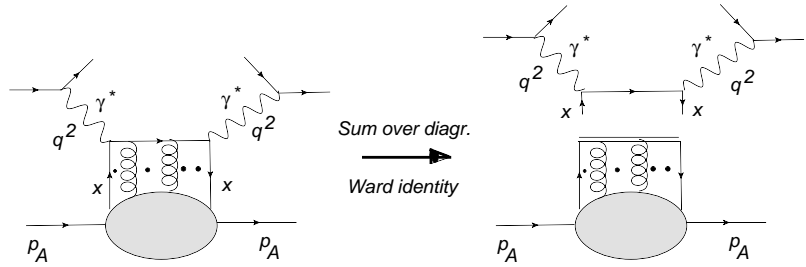}
\end{center}
\caption{Collinear (wrt  the target momentum $p_A$) gluonic exchanges  forming Wilson line of PDF }
\label{figDISnew2WilsonUncut}
\end{figure}

\noindent
{\bf 1. QCD factorisation and DIS.} The deep-inelastic scattering (DIS) is the simplest and most studied inclusive process in which the QCD factorization holds and in fact leads to description of DIS within the QCD improved parton model, see Fig.~\ref{figDISnew2amplhard} illustrating contribution of quark exchanges.  
%%%%%%%%%%%%%%%%%%%%%%%%%%%%%%%%%%%%%%%%%%%%%%%%%
%%%%%%%%%%%%%%%%%%%%%%%%%%%%%%%%%%%%%%%%%%%%%%%%%%%
The cross section of DIS $\sigma$ is a convolution in the longitudinal fractions $x$ of parton momenta and for the factorization scale $\mu$ of the cross-section $\sigma_j(x,\mu)$ for the process occurring on parton $j$,  with the forward parton $j$ distribution function (PDF)  $f_{j/h}(x,\mu)$
\begin{equation}
{\sigma}\! \sim \!\int\limits_{-1}^1\;\sigma_j(x,\mu)\,f_{j/h}(x,\mu)\ ,\;\;\;\;f_{j/h}(x,\mu)\! \sim \! \int dz^-\;e^{-ixp^+z^-} \;\langle p_A| \bar q(0,z^-,0_T)\,\Gamma\,W(z^-,0)\,q(0)\,|p_A\rangle\,.
\label{amplDIS}
\end{equation}
The PDF $f_{j/h}(x,\mu)$ is a forward matrix element of a non-local operator; its gauge invariance is assured by the appearance of the Wilson line 
\begin{equation}
W(z^-,0)= \mbox{P}\, e^{-ig\int\limits_0^{z^-} \, dy^- A^+(0,y^-,0_T)}\,,\; \;\;\;\;A^+ = A^{a\,+}t^a\,, \;\;\;\;\;W(z^-,0) = W(\infty^-,z^-)^\dagger\,W(\infty^-,0)
\label{WilsonL}
\end{equation}
emerging from exchanges of collinear gluons between the target and the hard CF (see Fig.~\ref{figDISnew2WilsonUncut}).
With the help of Ward identities one can convince oneself, that these exchanges effectively attach phases to the quark operators defining PDF, which means that the Wilson line $W(z^-,0)$ is in fact a product of two Wilson lines extended to infinity along the light-cone direction.
%$W(z^-,0) = W(\infty^-,z^-)^\dagger\,W(\infty^-,0)$.

\begin{figure}[h!]
\begin{center}
\epsfxsize=0.3\textwidth
\epsffile{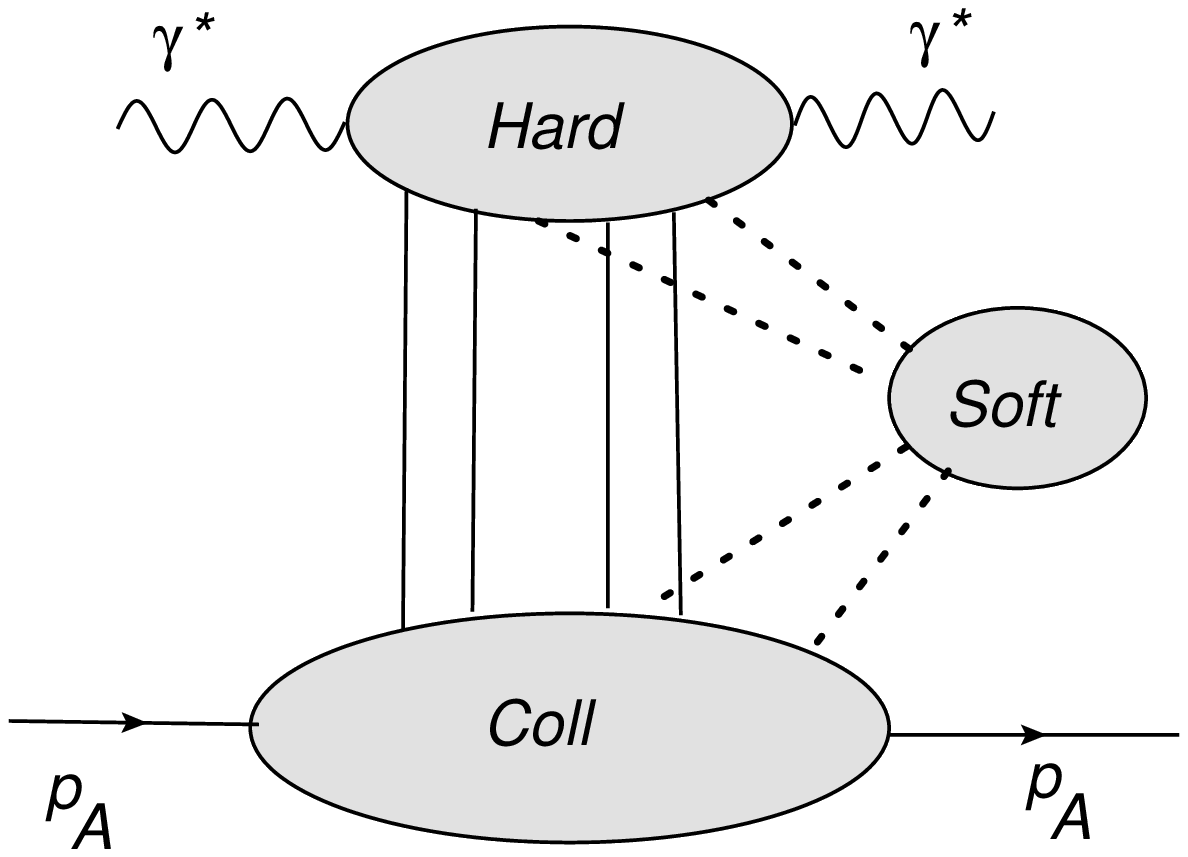}
\hspace*{2cm}
\epsfxsize=0.3\textwidth
\epsffile{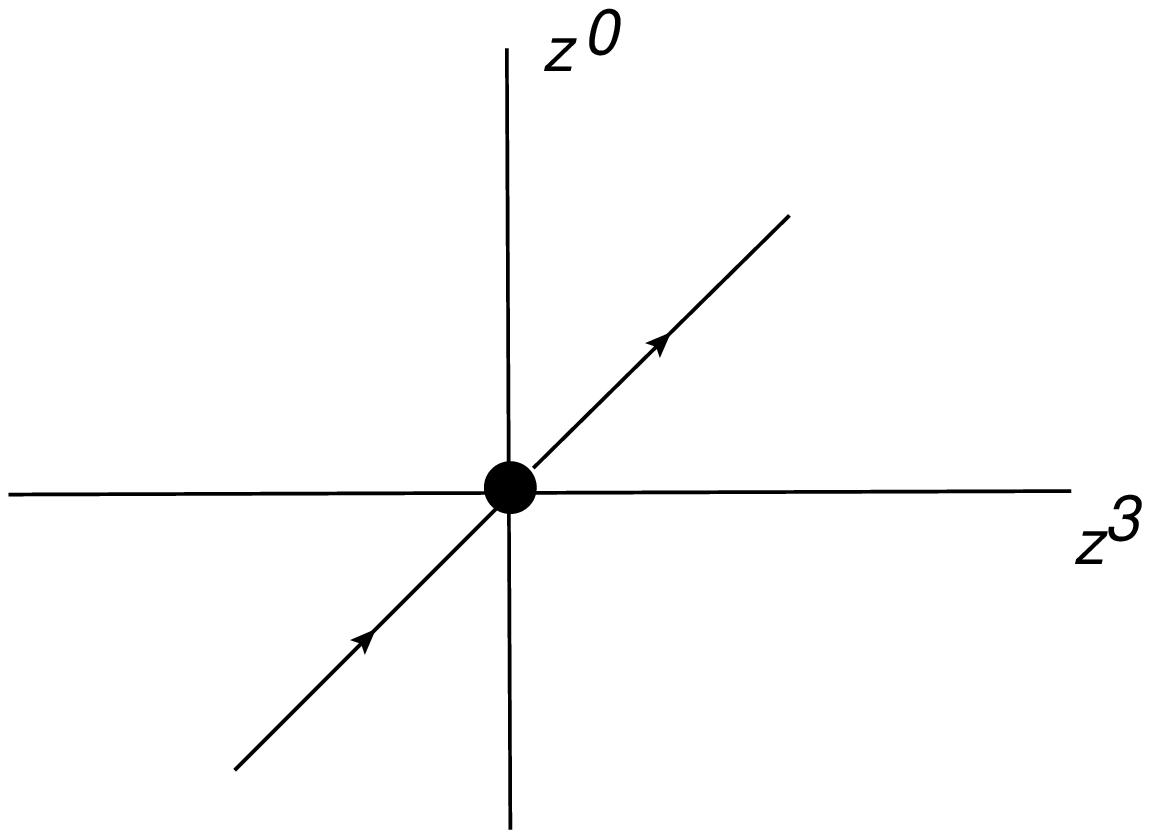}
\end{center}
\caption{DIS: an example of  a reduced diagram (left) together with its  space time diagram (right); solid lines are hard collinear (wrt  the target momentum $p_A$) partons, dashed lines are soft partonic lines, the point in the right figure corresponds to the reduced hard vertex; the nonlocalizable soft block  is not shown. }
\label{figDISRegionGood}
\end{figure}

\noindent
The mathematical tools and the basic steps of the proof of the factorization formula (\ref{amplDIS}) are based on an all-order analysis of Feynman diagrams in a peculiar kinematics, as reviewed, e.g. in \cite{CSSreview,Collinsbook} : 

\noindent
$\bullet$ Determination of reduced graphs with the help of the hard scale power counting method of Ref.~\cite{LibbySterman78}, which are built from the hard vertex ( a reduced diagram involving shrunk to the point propagators with a flow of the hard scale), the collinear block (a reduced diagram with hard longitudinal momenta) and the soft blocks 
(reduced diagram with only soft partons) connected either by soft or collinear lines. 
Fig.~\ref{figDISRegionGood} shows an example of the reduced diagram together with its the space time diagram
in the case of DIS. 

\begin{figure}[h!]
\begin{center}
\epsfxsize=0.9\textwidth
\epsffile{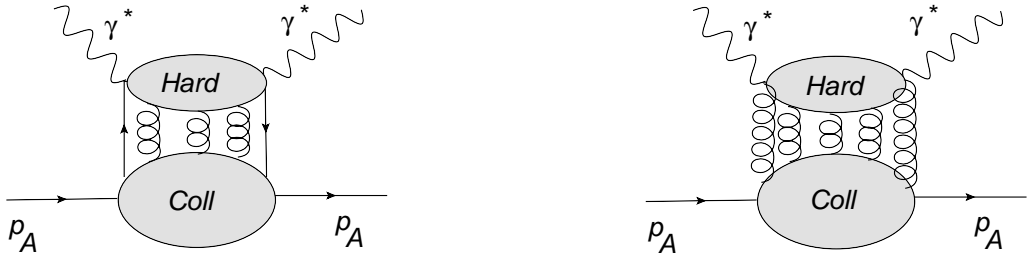}
\end{center}
\caption{DIS: the reduced diagram with the leading regions for quark (left) and gluon (right) exchanges; they involve only collinear blocks connected with the hard vertex by means of an arbitrary number of collinear gluon exchanges. }
\label{figDISRegionLeadingTwo}
\end{figure}

\noindent
$\bullet$ Separation from the reduced diagrams those with  the  leading regions in a way inspired by the method of Grammer-Yennie \cite{Grammeryennie73}. For DIS the reduced diagram with the leading regions is shown in Fig.~\ref{figDISRegionLeadingTwo}. In contrast to  Fig.~\ref{figDISRegionGood}, it does not contain the soft block  and includes only collinear gluon exchanges. 

\noindent
$\bullet$ The reduced diagrams with the leading regions contain in general an arbitrary power of the strong $\alpha_s$ and loops. 
The removal of associated with them UV divergences is performed by renormalisation of field operators defining GPDs.
The basic tool is the renormalisation group equation which governs the factorisation scale $\mu$ dependence of   $\sigma_j(x,\mu)$ and PDF $f_{j/h}(x,\mu)$. In the case of DIS, it is the well-known DGLAP equation  
\begin{equation}
\frac{d}{d\,\log \mu^2} \,f_{i/H}(x,\mu) = \sum\limits_j \int \frac{dy}{y}\, P_{i\,j}(y)\,f_{j/H}(\frac{x}{y},\mu)\;.
\label{EvolEq}
\end{equation}

\begin{figure}[h!]
\begin{center}
\epsfxsize=0.3\textwidth
\epsffile{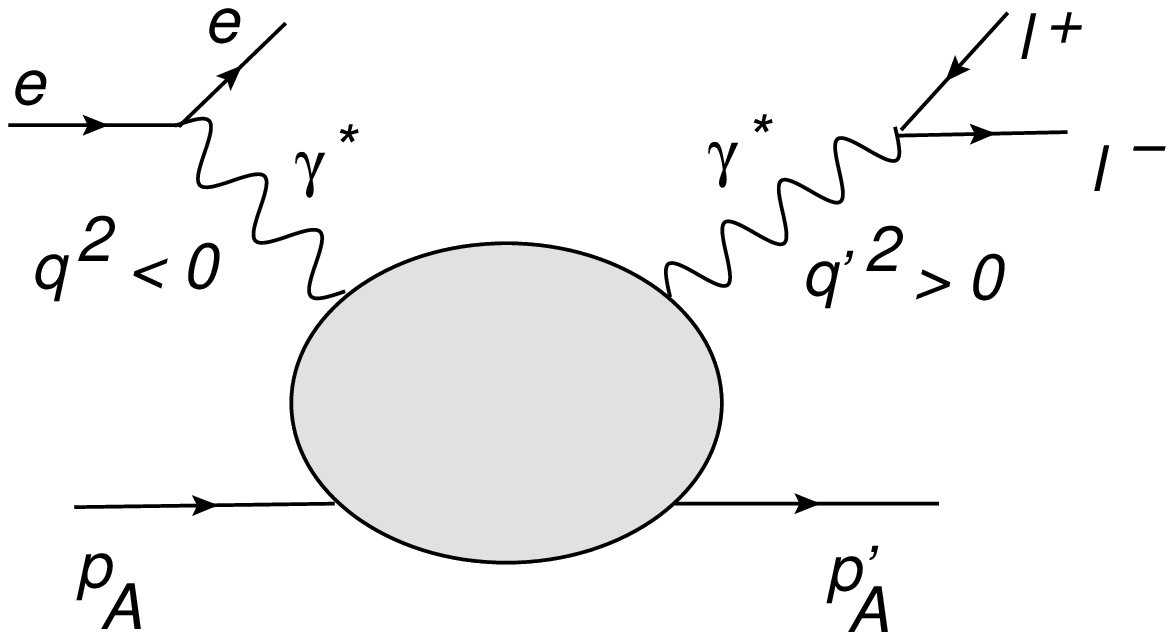}
\hspace*{0.5cm}
\epsfxsize=0.24\textwidth
\epsffile{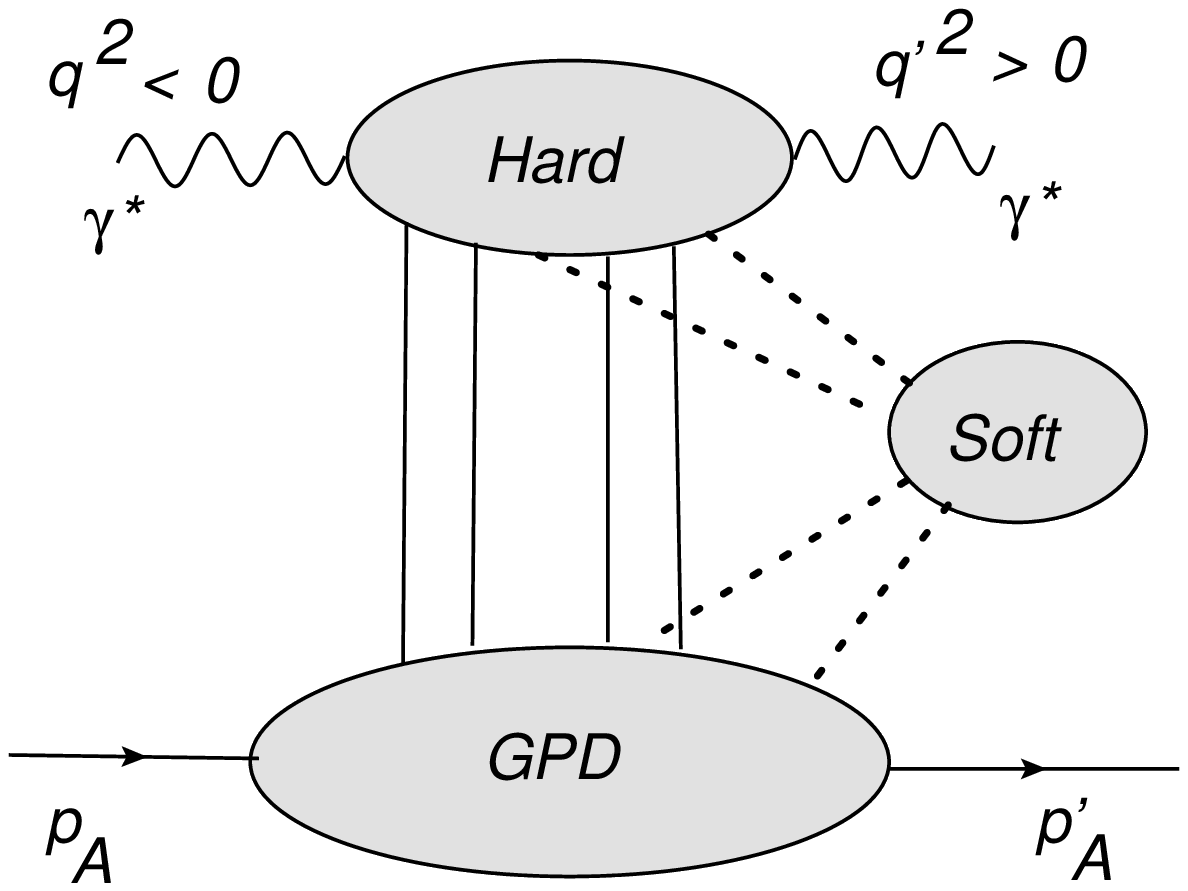}
\hspace*{0.5cm}
\epsfxsize=0.3\textwidth
\epsffile{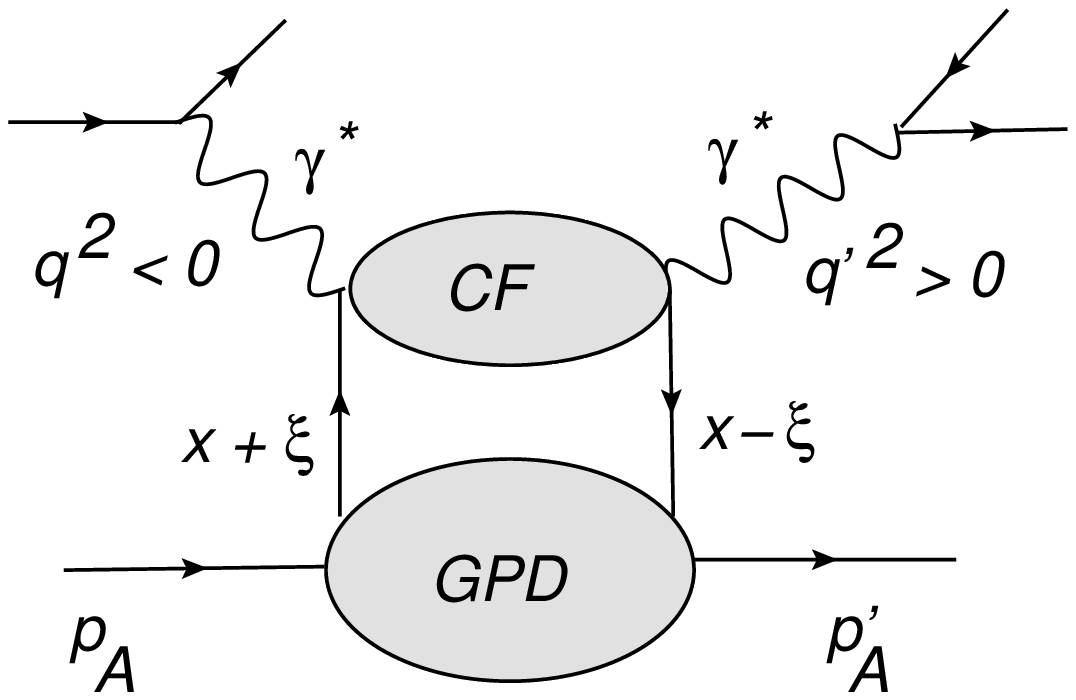}
\end{center}
$\hspace*{2.2cm}(a) \hspace*{4.5cm} (b) \hspace*{4.3cm} (c)$
\caption{ (a) - the  DDVCS process; (b) - example of a reduced diagram; (c) - the factorized form of scattering amplitude}
\label{figDDVCS}
\end{figure}

\noindent
{\bf 2. QCD factorization in DDVCS.} 
Double deeply virtual Compton scattering (DDVCS) is the exclusive process 
$e\; N \rightarrow e' \; N'\;l^+\;l^-$
illustrated in Fig.~\ref{figDDVCS}a. It involves one space-like incoming virtual photon and the outgoing one being timelike.  Because of its close similarity to DIS, the reduced regions (see Fig.~\ref{figDDVCS}b) are the same for both processes. Consequently, the proof of factorization in DIS can be extended directly to DDVCS. 
%\begin{equation}
%e\; N \rightarrow e' \; N'\;l^+\;l^-
%\label{reacDDVCS}
%\end{equation}
The essential new (kinematical) feature of DDVCS with respect to DIS is a non-zero momentum transfer in the t-channel leading to nonvanishing skewness $\xi =\frac{p^+_A- {p'}^+_A}{p^+_A + {p'}^+_A}$. It results in the form of the factorization formula for the scattering amplitude in  Fig.~\ref{figDDVCS}c involving quark correlator
\begin{equation}
{\cal A}(\xi) \sim \int\limits_{-1}^1\;H(x,\xi,\mu)\,f_{j/h}(x,\xi,\mu)\;,
\label{amplDDVCS}
\end{equation}
in which dependence on $\xi$ appears both in the hard CF $H(x,\xi,\mu)$ and in
 the non-forward generalized parton distribution (GPD) $f_{j/h}(x,\xi,\mu)$   \cite{Mueller94}, \cite{Rad96}, \cite{Ji96}
\begin{equation}
\hspace*{-0.8cm}
f_{j/h}(x,\xi,\mu) \sim \int dz^-\;e^{ixP^+z^-} \;\langle {p'}_A| \bar q(0,-\frac{1}{2}z^-,0_T)\,\Gamma\,W(-\frac{1}{2}z^-,\frac{1}{2}z^-)\,q(0, \frac{1}{2}z^-,0_T)\,|p_A\rangle \;.
\label{defGPD}
\end{equation}
As in the DIS case, the presence of the Wilson line $W(-\frac{1}{2}z^-,\frac{1}{2}z^-)$ in (\ref{defGPD}) guarantees
the gauge invariance of the GPD $f_{j/h}(x,\xi,\mu)$.
%\begin{equation}
%W(-\frac{1}{2}z^-,\frac{1}{2}z^-)= \mbox{P}\, e^{ig\int\limits_{1/2z^-}^{-1/2z^-} \, dy^- A^+(0,y^-,0_T)} \;\;\;\;\;\;\;A^+ = A^{a\,+}t^a
%\label{WLineGPD}
%\end{equation}

\begin{figure}[h!]
\begin{center}
\epsfxsize=0.25\textwidth
\epsffile{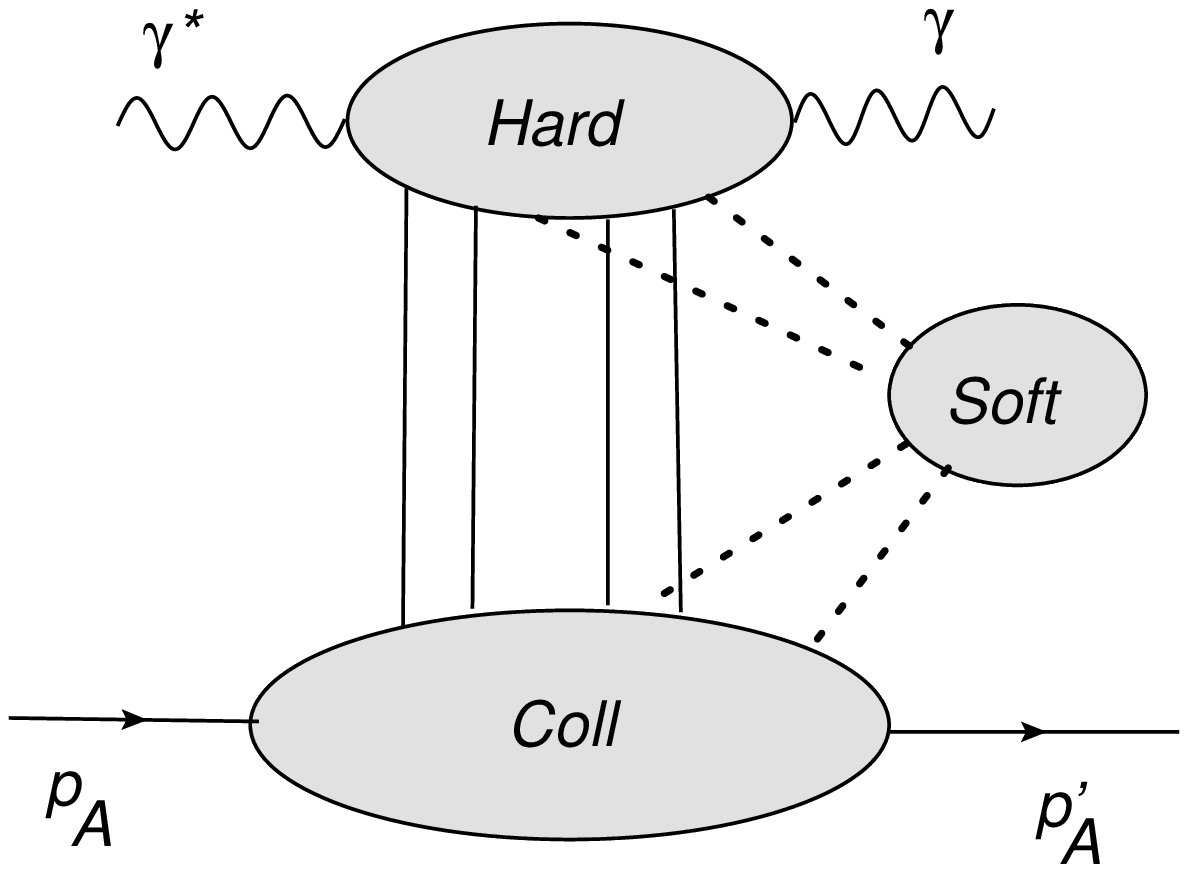}
\hspace*{0.5cm}
\epsfxsize=0.3\textwidth
\epsffile{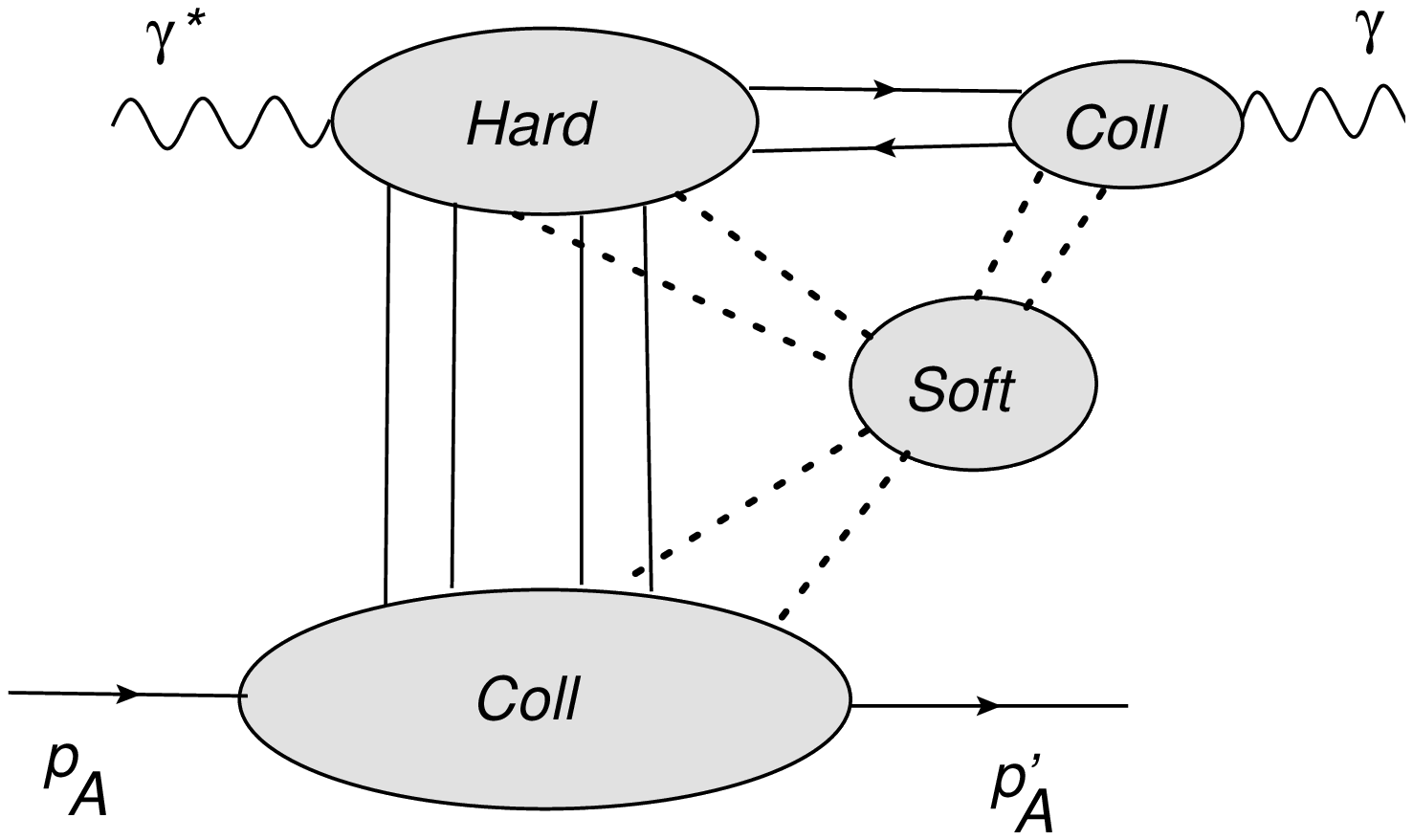}
\hspace*{0.5cm}
\epsfxsize=0.3\textwidth
\epsffile{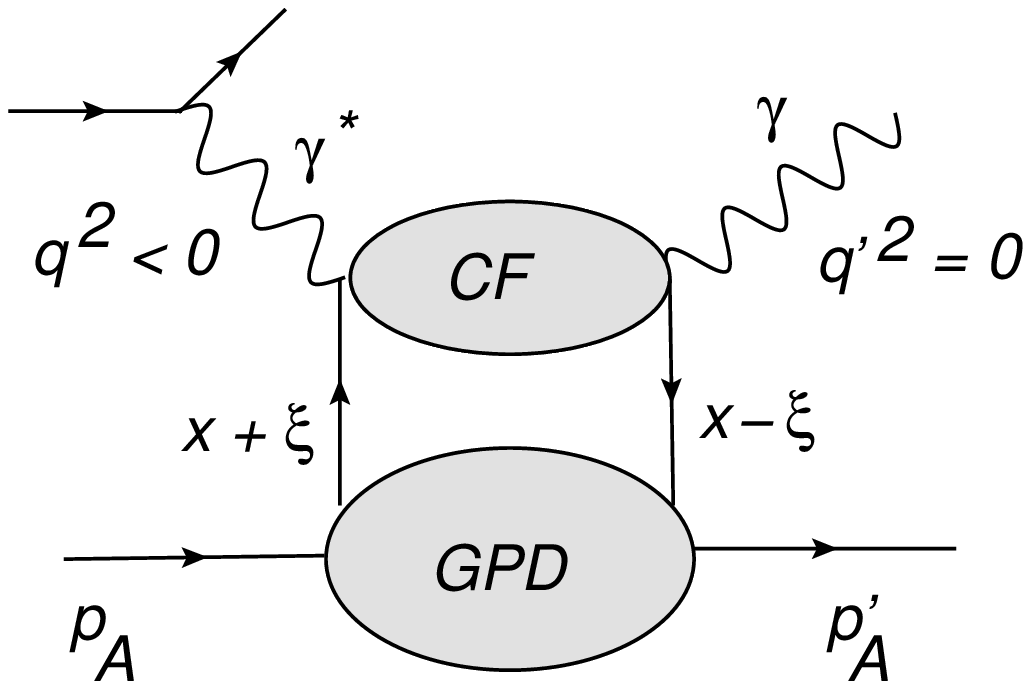}
\end{center}
$\hspace*{2.2cm}(a) \hspace*{4.5cm} (b) \hspace*{4.5cm} (c)$
\caption{DVCS: (a,b) - examples of reduced diagrams; (c) - the factorized form of scattering amplitude}
\label{figDVCSregion}
\end{figure}

\noindent
{\bf 3. QCD factorisation in DVCS.}
Deeply virtual Compton scattering (DVCS) occurs when the outgoing photon in DDVCS is a real one, i.e.
$e\; N \rightarrow e' \; N'\;\gamma$. The lack of the hard scale related to virtuality of the final $\gamma$ requires to consider not only the reduced diagram of Fig.~\ref{figDVCSregion}a discussed already, but also the one shown in
Fig.~\ref{figDVCSregion}b in which real $\gamma$ couples to additional collinear block.
However,  due to the point-like coupling of $\gamma$ to quarks, the diagram
in Fig.~\ref{figDVCSregion}b gives a non-leading contribution. Thus the factorisation theorem for DVCS has a form given by Eqs.~(\ref{amplDDVCS}, \ref{defGPD}), like for the DDVCS process \cite{Mueller94}, \cite{Rad96}, \cite{Ji96}.
The DVCS process $e\; N \rightarrow e' \; N'\;\gamma$ or  its  crossed version named TCS for   time-like Compton scattering  $\gamma N \rightarrow l^+ \;l^- \; N'$ with a  lepton pair $l^+ \;l^-$ \cite{TCS} are of big phenomenological importance due to experiments performed or planned at JLab, Hermes, RHIC, LHC. For a review of  phenomenological aspects of these processes I refer to the talks at this conference presented by F. Sabatie \cite{Sabatie} and   J. Wagner \cite{Wagner}.

\begin{figure}[h!]
\begin{center}
\epsfxsize=0.75\textwidth
\epsffile{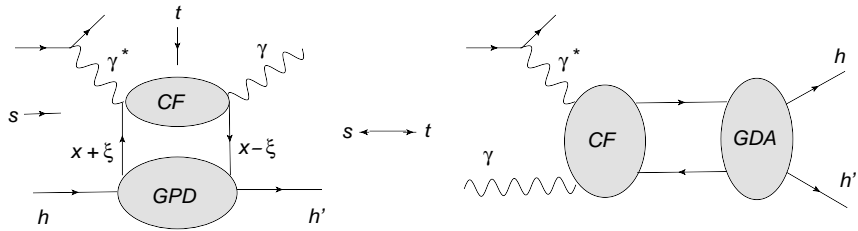}
\end{center}
\caption{The relationship between DVCS and $\gamma^*$ $\gamma$ scattering process due to the  $s\;\leftrightarrow\;t$ channels crossing.}
\label{figDVCStoGDA}
\end{figure}

\noindent
{\bf 4. QCD factorization with GDAs.} The generalized distribution amplitudes (GDAs) are non-perturbative, soft
matrix elements of non-local operator (as the quark one appearing in the definition of GPD (\ref{defGPD})), between a vacuum and a state of  (at least two) produced hadrons  \cite{Mueller94}, \cite{Diehl98}, \cite{Pol99}, \cite{Kivel99}. GDAs are building blocks in factorized scattering amplitude for e.g. phenomenologically important  $\gamma^*$ $\gamma$ scattering processes studied experimentally by BABAR and BELLE collaborations. The $\gamma^*$ $\gamma$ scattering process can be viewed (see Fig.~\ref{figDVCStoGDA}) as a $t$-channel crossed version of the DVCS. 
The kinematical condition 
$s \ll |t| \sim Q^2$
between
the total $\gamma^*\,\gamma$ energy square, $s$,  the square of the momentum transfer $t$ between $\gamma^*\; h$ and the virtuality $Q$ of $\gamma^*$ 
 ensures for this process a  validity of similar arguments of the QCD factorization like in the DVCS case, including an impact representation of hadronization \cite{PSGDA}.

\begin{figure}[h!]
\begin{center}
\epsfxsize=0.25\textwidth
\epsffile{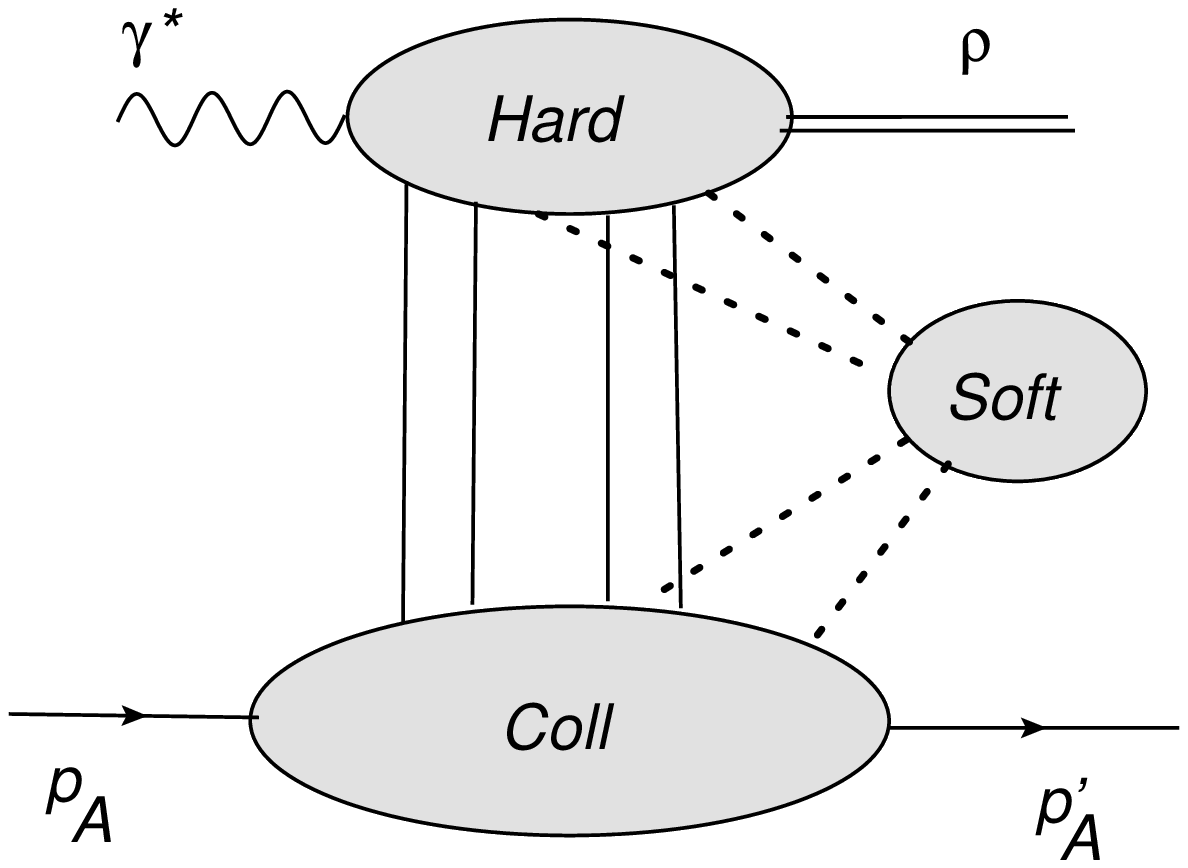}
\hspace*{0.2cm}
\epsfxsize=0.30\textwidth
\epsffile{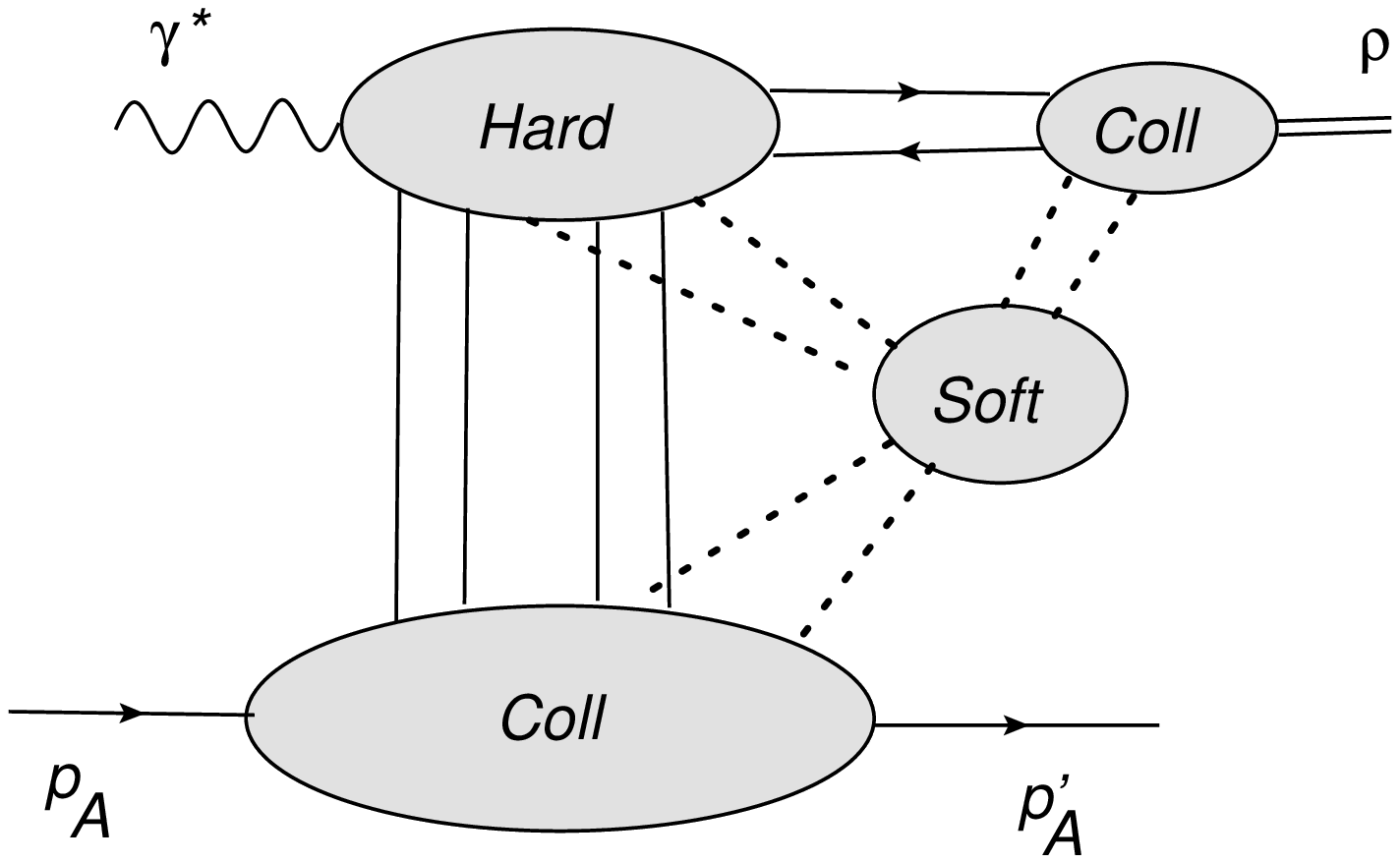}
\hspace*{0.2cm}
 \epsfxsize=0.39\textwidth
\epsffile{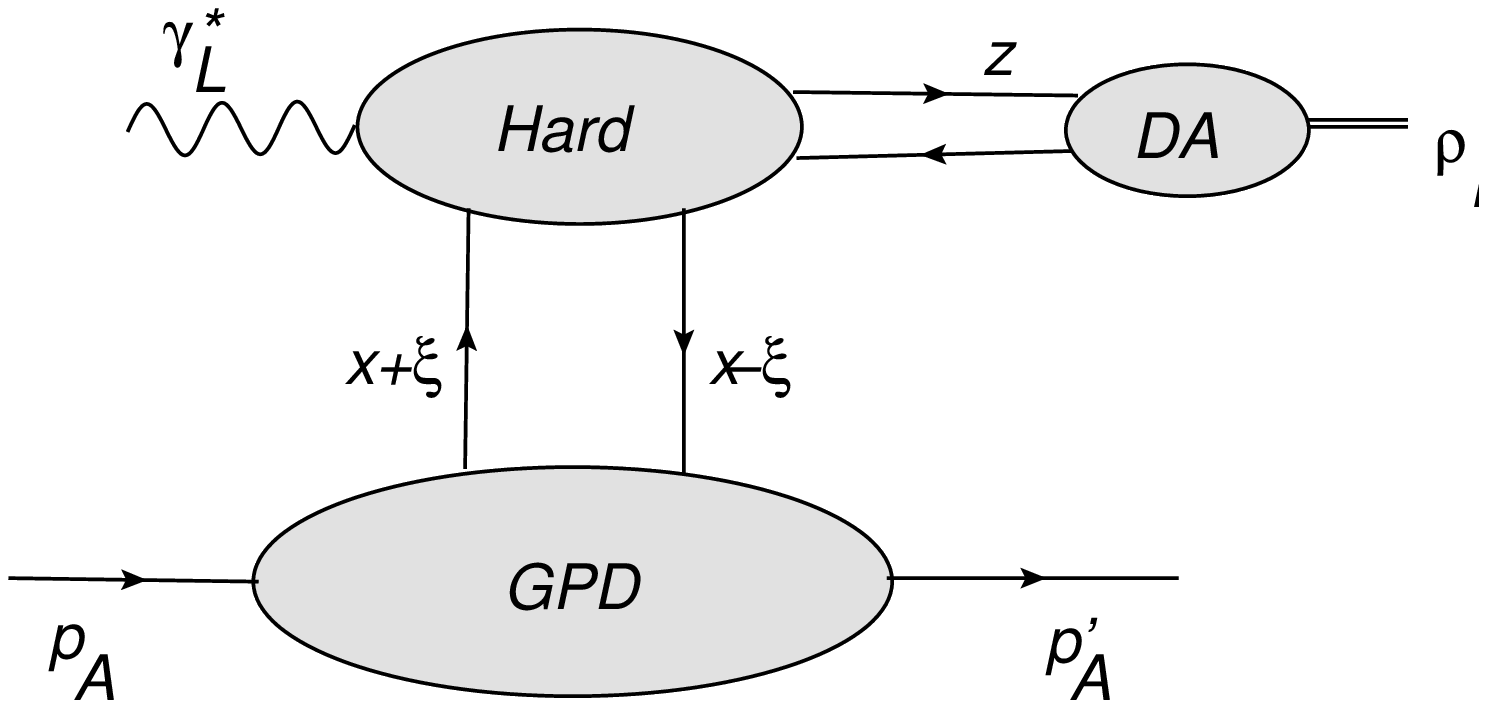}
\end{center}
$\hspace*{1.5cm}(a) \hspace*{4.1cm} (b) \hspace*{4.5cm} (c)$
\caption{$\rho$-meson production: (a,b) - examples of reduced diagrams; (c) - the factorized form of scattering amplitude in the case of $\rho_L$-meson production}
\label{figrhoregion}
\end{figure}

\begin{figure}[h!]
\begin{center}
 \epsfxsize=0.9\textwidth
\epsffile{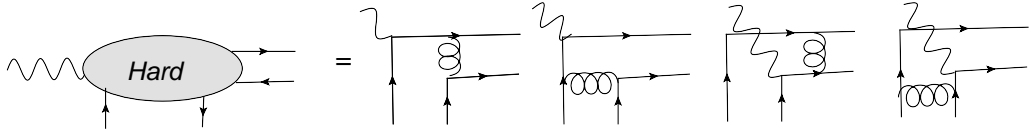}
\end{center}
\caption{Feynman diagrams contributing in Born approximation to the hard CF in $\rho_L$ production}
\label{figRhoHard}
\end{figure}

\noindent
{\bf 5. QCD factorization in meson production.}
For definiteness, let us concentrate on the $\rho$-meson production in $\gamma^*$-nucleon scattering. The analysis of conditions for QCD factorization proceeds along the same lines like in the case of DVCS and leads to the similar reduced diagrams shown in   Figs.~\ref{figrhoregion} a,b
The crucial difference in QCD factorization in DVCS and in a meson production is due to the fact that - contrary to the photon -  a meson is a bound state of partons which doesn't have a point-like coupling to quarks. As a result, the role of the soft block in   Fig.~\ref{figrhoregion}b  which can spoil factorization depends on the polarization state of the produced $\rho$-meson. It turns out \cite{Brodskyetal94}, \cite{Collinsetal97}, that  the QCD factorization holds  for the production of longitudinally polarized $\rho_L$-meson  and has the form shown in Fig.~\ref{figrhoregion}c 
\begin{equation}
{\cal A}(\xi,t) \sim \int\limits_{-1}^{1}dx \;\int\limits_0^1dz\;H(x,z,\mu)\,\Phi^q(z,\mu)\,f_{q/h}(x,\xi,t,\mu)\;.
\label{ScattAmplMeson}
\end{equation}
The $\rho_L$-meson is described in the factorized scattering amplitude (\ref{ScattAmplMeson}) by the twist-2 distribution amplitude (DA) $\Phi^q(z,\mu)$, a nonperturbative, integrated over transverse partonic momenta wave function \cite{Balletal96}
\begin{equation}
\Phi^q(z,\mu) \sim \int dz^-\; e^{i(2z-1)p^+x^-} \langle \rho_L(p)|\bar q(-\frac{x}{2})W(-\frac{x}{2},\frac{x}{2})\gamma^+q(\frac{x}{2})|0\rangle|_{x^+=0={\bf x}} \;, 
\label{DAmeson}
\end{equation}
which convoluted with
 the hard CF  $H(x,z,\mu)$ receiving (in the Born approximation) contributions from the diagrams in Fig.~\ref{figRhoHard} and the GPD $f_{q/h}(x,\xi,t,\mu)$ gives a dominant (in powers of the hard scale $Q$) contribution to the scattering amplitude ${\cal A}(\xi,t)$.
It is also important to emphasize, that in the factorized expression (\ref{ScattAmplMeson}) for $\rho_L$-meson production there appears the same  GPD
as in Eq.~(\ref{amplDDVCS}) for DDVCS and DVCS processes, showing universality property of GPDs within the QCD collinear factorization.

\begin{figure}[h!]
\begin{center}
\epsfxsize=0.52\textwidth
\epsffile{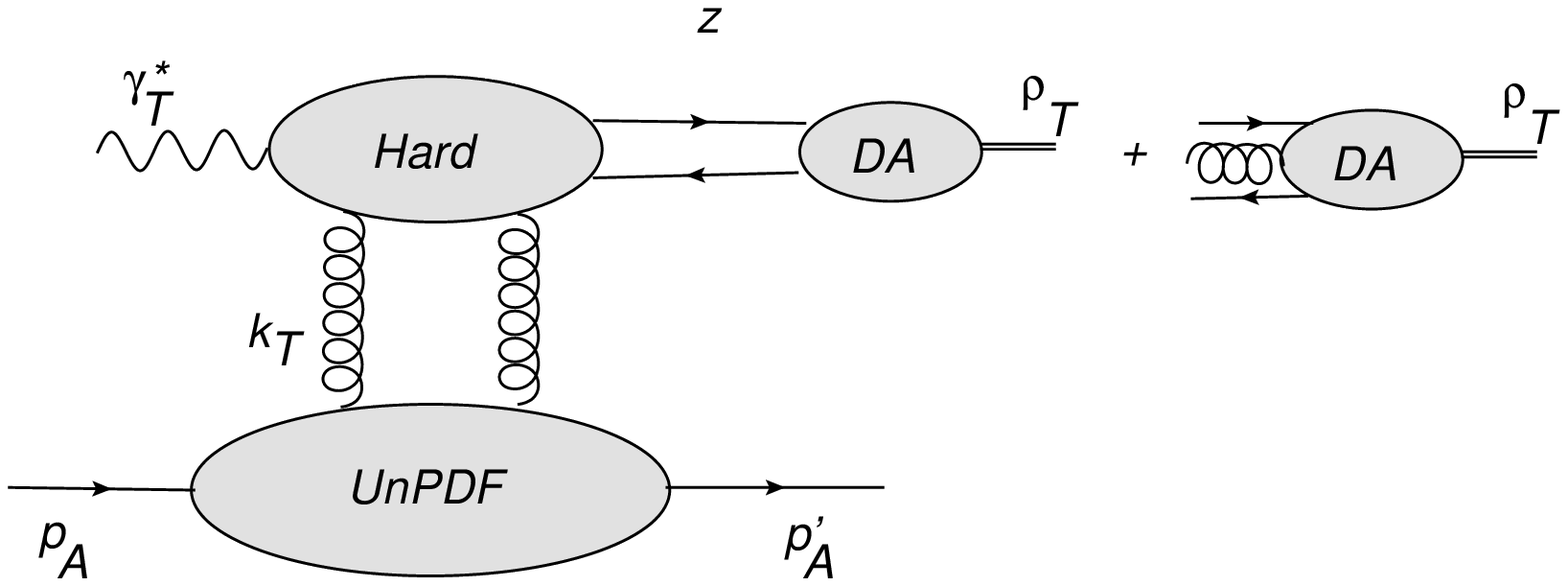}
\hspace*{0.4cm}
\epsfxsize=0.38\textwidth
\epsffile{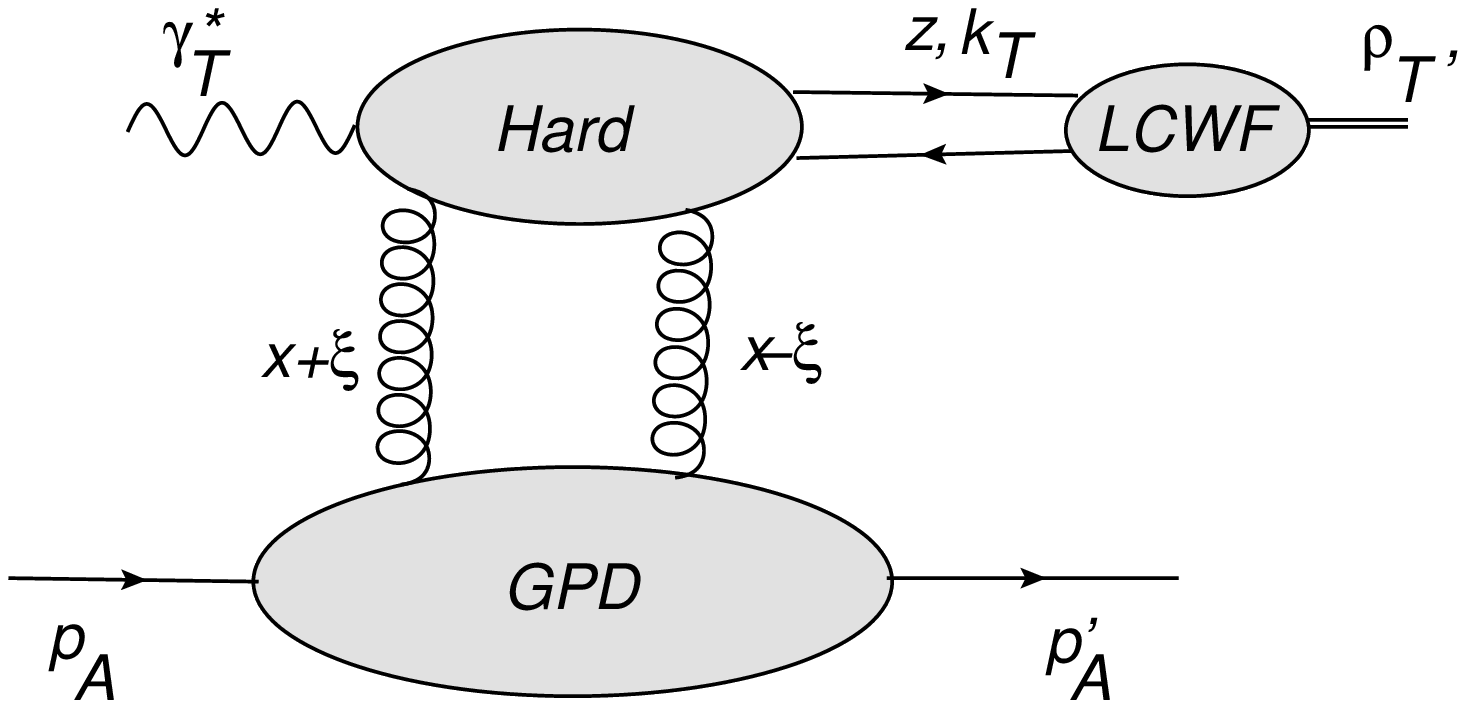}
\end{center}
$\hspace*{2.5cm}(a) \hspace*{8.1cm} (b) $
\caption{Two approaches to the hard electroproduction of $\rho_T$: (a) with higher Fock states in $\rho_T$ and $k_T$ momenta in $t$-channel; (b) with the LCWF of $\rho_T$. }
\label{figRhoFactor}

\end{figure}

\noindent
For production of transversally polarized $\rho_T$-meson the QCD (collinear) factorization is violated \cite{Mankiewiczetal99}, as the convolution integrals in (\ref{ScattAmplMeson}) diverge at the edge of kinematical domain and lead to the appearance of  end-point singularities. It is due to the fact, that production of $\rho_T$-meson  is related to a non-leading asymptotic in powers of $Q$, which is described by the DA having higher twist-3. In this case the role of transverse partonic momenta increases. There are two attempts to overcome the above problem, especially for processes occurring at high energies within a small-$x$ kinematics. The attempt proposed in  
\cite{Anikin01},     \cite{Anikin03},    \cite{Anikinetal10},    \cite{Besseetal11} consists in including  in the description of $\rho_T$ not only the DA defined by the  lowest (quark) Fock state but also by the higher Fock state involving gluons, and keeping nonvanishing transverse momenta of $t$-channel partons (the factorization $k_T$), see Fig.~\ref{figRhoFactor}a. More details of this approach is presented in the talk by A. Besse at this conference \cite{talkBesse}.  

\noindent
The attempt proposed in \cite{Gol05} consists in description of a $\rho$-meson in terms of the light-cone wave function (LCWF) in which
 partons have nonvanishing transverse momenta. The partons in the $t$-channel are kept collinear, see Fig.~\ref{figRhoFactor}b. The inclusion in the hard part of Sudakov form-factor permits to avoid  end-point singularities. For more details on this approach I refer to the talk by P. Kroll  at this conference \cite{talkKroll}.

\begin{figure}[h!]
\begin{center}
\epsfxsize=0.33\textwidth
\epsffile{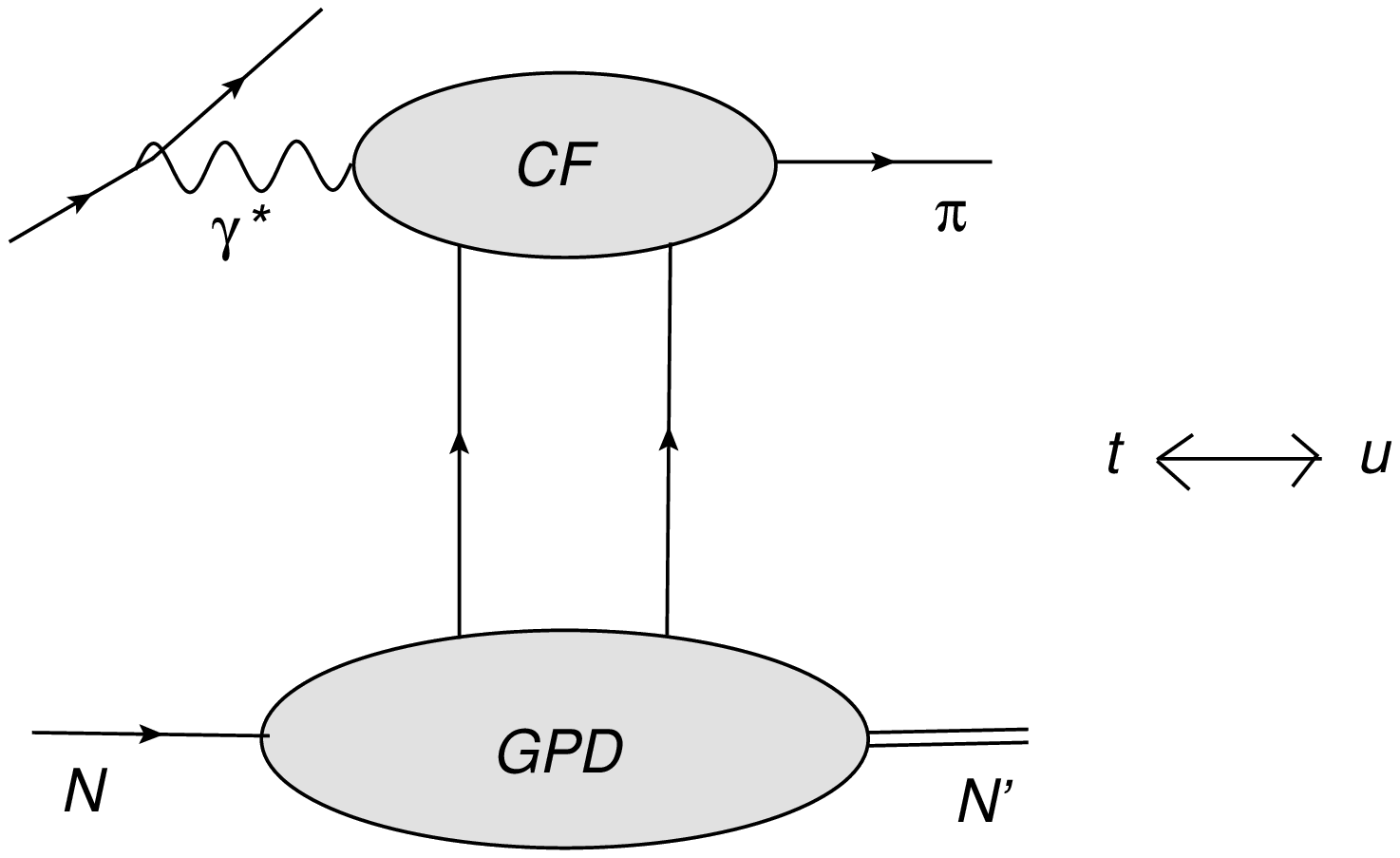}
\hspace*{0.5cm}
\epsfxsize=0.25\textwidth
\epsffile{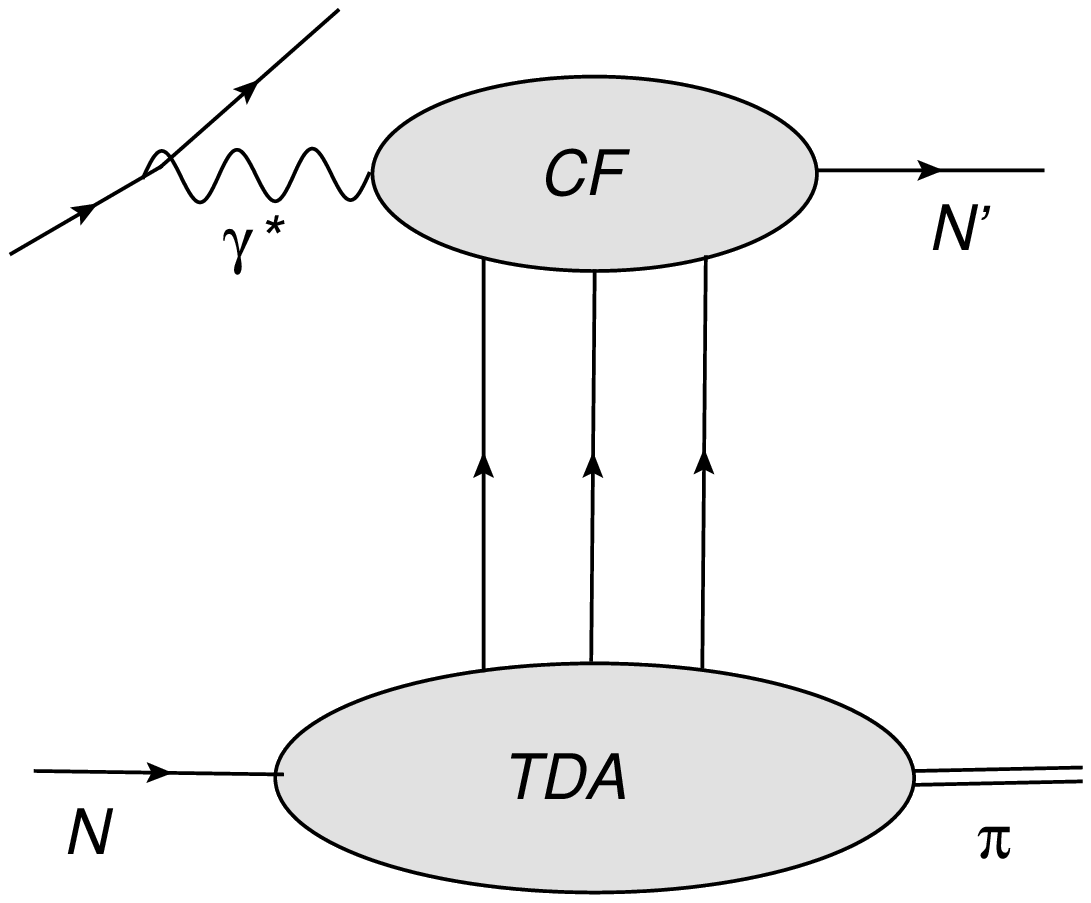}
\end{center}
\caption{The relationship of forward and backward meson production due crossing between $s$ and $u$ channels}
\label{figTDA}
\end{figure}

\noindent
{\bf 6. Extension of factorization to backward meson production.} Inspired by the $s \leftrightarrow t$ channel crossing relating DVCS with $\gamma^* \gamma$ scattering process in Sec.6, one can apply a similar procedure in the case of meson $M$ production process $\gamma^*\; N\;\to\;M\;N'$ discussed above. The $s \leftrightarrow u$ channel crossed version of this process corresponds to the backward meson $M$ scattering $\gamma^*\; N\;\to\;N'\;M$, see Fig.~\ref{figTDA}, in which the  outgoing meson $\pi$ flows almost in direction of  incoming nucleon $N$, i.e.  opposite to the direction of flow of incoming $\gamma^*$.   

\begin{figure}[h!]
\begin{center}
\epsfxsize=0.70\textwidth
\epsffile{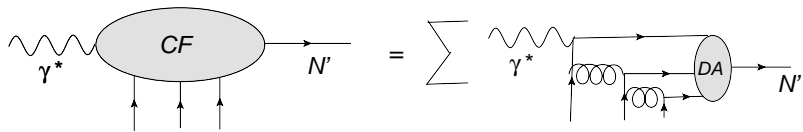}
\end{center}
\caption{The hard part for backward meson production in Born approximation.}
\label{figDVCStoTDAhardpart}
\end{figure}

\noindent
It is  natural to expect that the factorization theorem involves a new non-perturbative matrix element of non-local operator between states of initial nucleon $N$ and final meson $M$, called the nucleon-meson transition distribution amplitude (TDA) \cite{TDA}. The conservation of baryonic number requires that the non-local operator carries baryon quantum numbers, i.e. it is given by a non-local product of three quark operators. Thus the hard part of backward meson production in the Born approximation contains diagrams shown in Fig.~\ref{figDVCStoTDAhardpart}
convoluted with the baryon DA.
Since the hard part in Fig.~\ref{figDVCStoTDAhardpart} is of higher order in strong coupling constant than the CF for forward meson production in Fig.~\ref{figRhoHard}
one expects smaller cross section for the backward meson production than for its forward partner. This makes phenomenology with TDA quite challenging. The experimental studies with TDAs in $\gamma^*\;N$ scattering are planned at JLab whereas the $\bar N \,N \to \gamma^* \, \pi$ process will be studied at Panda (GSI).
For more details on the phenomenology of TDAs, I refer to the talk by K. Semenov-Tian-Shansky at this conference \cite{talkKirill}.

\noindent
{\bf 7. Further developments.} The above review is  by no means exhaustive. It does not include the important extension of QCD factorization to  semi-inclusive DIS processes (SIDIS) in which final state one observes a hadron with nonvanishing $k_T$, or to the Drell-Yan  process with  produced virtual $\gamma^*$ having nonvanishing $k_T$. The description of these processes within  QCD factorization requires modification of their factorization formulas (like Eq.~(\ref{amplDIS}) for DIS) by including effects of transverse parton momenta both in their hard CFs and in parton distribution functions, which are replaced by the transverse momentum dependent PDFs (TMDs).   
Since the non-local operators defining TMDs are separated also in  transverse directions they involve more complicated Wilson lines as those in (\ref{WilsonL}) in DIS pointed only in  the light-like direction.
Consequently,
the factorization with TMDs requires  redefinition of the notion of universality of TMDs which start to be process dependent.  This leads to concrete phenomenological predictions which can be checked in 
 experiment. These phenomenological issues with TMDs are discussed in the talks by M. Anselmino \cite{talkAnselmino} and by L. Gamberg \cite{talkGamberg}.


\begin{thebibliography}{99}

\bibitem{CSSreview}J.~C.~Collins, D.~E.~Soper and G.~F.~Sterman,
  %``Factorization of Hard Processes in QCD,''
  Adv.\ Ser.\ Direct.\ High Energy Phys.\  {\bf 5} (1988) 1
\bibitem{Collinsbook} J. Collins, "Foundations of Perturbative QCD", Cambridge University Press 2011.

\bibitem{LibbySterman78} S.~B.~Libby and G.~F.~Sterman,
  %``Jet and Lepton Pair Production in High-Energy Lepton-Hadron and Hadron-Hadron Scattering,''
  Phys.\ Rev.\ D {\bf 18} (1978) 3252; {\it ibid} Phys.\ Rev.\ D {\bf 18} (1978) 4737.
  
  \bibitem{Grammeryennie73}G.~Grammer, Jr. and D.~R.~Yennie,
  %``Improved treatment for the infrared divergence problem in quantum electrodynamics,''
  Phys.\ Rev.\ D {\bf 8} (1973) 4332.
  
  \bibitem{Mueller94} D.~M{\"u}ller, D.~Robaschik, B.~Geyer, F.~M.~Dittes and J.~Horejsi,
  %``Wave functions, evolution equations and evolution kernels from light ray operators of QCD,''
  Fortsch.\ Phys.\  {\bf 42} (1994) 101.
  
  \bibitem{Rad96} A.~V.~Radyushkin,
  %``Asymmetric gluon distributions and hard diffractive electroproduction,''
  Phys.\ Lett.\ B {\bf 385} (1996) 333.
  
  \bibitem{Ji96} X.~-D.~Ji,
  %``Deeply virtual Compton scattering,''
  Phys.\ Rev.\ D {\bf 55} (1997) 7114.

\bibitem{TCS}
E.~R.~Berger, M.~Diehl and B.~Pire,
  %``Timelike Compton scattering: Exclusive photoproduction of lepton pairs,''
  Eur.\ Phys.\ J.\  C {\bf 23}, 675 (2002);
  %%CITATION = EPHJA,C23,675;%%
 B.~Pire, L.~Szymanowski and J.~Wagner,
  %``Can one measure timelike Compton scattering at LHC?,''
  Phys.\ Rev.\ D {\bf 79} (2009) 014010 and
   Phys.\ Rev.\ D {\bf 83}, 034009 (2011);
  %%CITATION = ARXIV:1101.0555;%%
  D.~Mueller, B.~Pire, L.~Szymanowski and J.~Wagner,
  %``On timelike and spacelike hard exclusive reactions,''
  arXiv:1203.4392 [hep-ph].
  %%CITATION = ARXIV:1203.4392;%%
   %%CITATION = ARXIV:0811.0321;%%

  \bibitem{Sabatie} F. Sabatie,  PoS QNP {\bf 2012}, 016 (2012)
  [arXiv:1207.4655 [hep-ex]].
  %%CITATION = ARXIV:1207.4655;%%
  
  \bibitem{Wagner} J. Wagner, talk at this conference,  arXiv:1207.2301 [hep-ph].
  %%CITATION = ARXIV:1207.2301;%%
  
  \bibitem{Diehl98}M.~Diehl, T.~Gousset, B.~Pire and O.~Teryaev,
  %``Probing partonic structure in gamma* gamma ---> pi pi near threshold,''
  Phys.\ Rev.\ Lett.\  {\bf 81} (1998) 1782; 
  M.~Diehl, T.~Gousset and B.~Pire,
  %``Exclusive production of pion pairs in gamma* gamma collisions at large Q**2,''
  Phys.\ Rev.\ D {\bf 62}, 073014 (2000).
  %%CITATION = HEP-PH/0003233;%%
  
  
  \bibitem{Pol99} M.~V.~Polyakov,
  %``Hard exclusive electroproduction of two pions and their resonances,''
  Nucl.\ Phys.\ B {\bf 555} (1999) 231.
  
  \bibitem{Kivel99} N.~Kivel, L.~Mankiewicz and M.~V.~Polyakov,
  %``NLO corrections and contribution of a tensor gluon operator to the process gamma* gamma ---> pi pi,''
  Phys.\ Lett.\ B {\bf 467} (1999) 263.
  
 
\bibitem{PSGDA} 
  B.~Pire and L.~Szymanowski,
  %``Impact representation of generalized distribution amplitudes,''
  Phys.\ Lett.\ B {\bf 556}, 129 (2003).
  %%CITATION = HEP-PH/0212296;%%
  
  \bibitem{Brodskyetal94} S.~J.~Brodsky, L.~Frankfurt, J.~F.~Gunion, A.~H.~Mueller and M.~Strikman,
  %``Diffractive leptoproduction of vector mesons in QCD,''
  Phys.\ Rev.\ D {\bf 50} (1994) 3134.
  
  \bibitem{Collinsetal97}J.~C.~Collins, L.~Frankfurt and M.~Strikman,
  %``Factorization for hard exclusive electroproduction of mesons in QCD,''
  Phys.\ Rev.\ D {\bf 56} (1997) 2982.
  
  \bibitem{Balletal96} P.~Ball and V.~M.~Braun,
  %``The Rho meson light cone distribution amplitudes of leading twist revisited,''
  Phys.\ Rev.\ D {\bf 54} (1996) 2182.
  
  \bibitem{Mankiewiczetal99} L.~Mankiewicz and G.~Piller,
  %``Comments on exclusive electroproduction of transversely polarized vector mesons,''
  Phys.\ Rev.\ D {\bf 61} (2000) 074013.
  
  \bibitem{Anikin01} I.~V.~Anikin and O.~V.~Teryaev,
  %``Wandzura-Wilczek approximation from generalized rotational invariance,''
  Phys.\ Lett.\ B {\bf 509} (2001) 95.
  
  \bibitem{Anikin03} I.~V.~Anikin and O.~V.~Teryaev,
  %``Genuine twist three in exclusive electroproduction of transversely polarized vector mesons,''
  Phys.\ Lett.\ B {\bf 554} (2003) 51.
  
  \bibitem{Anikinetal10} I.~V.~Anikin, D.~Yu.~Ivanov, B.~Pire, L.~Szymanowski and S.~Wallon,
  %``QCD factorization of exclusive processes beyond leading twist: gamma*T ---> rho(T) impact factor with twist three accuracy,''
  Nucl.\ Phys.\ B {\bf 828} (2010) 1; {\it ibid} 
  %``On the description of exclusive processes beyond the leading twist approximation,''
  Phys.\ Lett.\ B {\bf 682} (2010) 413.
 
  \bibitem{Besseetal11} I.~V.~Anikin, A.~Besse, D.~Yu~Ivanov, B.~Pire, L.~Szymanowski and S.~Wallon,
  %``A phenomenological study of helicity amplitudes of high energy exclusive leptoproduction of the rho meson,''
  Phys.\ Rev.\ D {\bf 84} (2011) 054004.
  
  \bibitem{Gol05} S.~V.~Goloskokov and P.~Kroll,
  %``Vector meson electroproduction at small Bjorken-x and generalized parton distributions,''
  Eur.\ Phys.\ J.\ C {\bf 42} (2005) 281.
  
  \bibitem{talkBesse} A. Besse, talk at this conference, arXiv:1207.2503 [hep-ph].
  %%CITATION = ARXIV:1207.2503;%%
  
  \bibitem{talkKroll} P. Kroll, talk at this conference, arXiv:1206.6215 [hep-ph].
  %%CITATION = ARXIV:1206.6215;%%
 
 \bibitem{TDA}
  L.~L.~Frankfurt, P.~V.~Pobylitsa, M.~V.~Polyakov and M.~Strikman,
  %``Hard exclusive pseudoscalar meson electroproduction and spin structure  of
  %a nucleon,''
  Phys.\ Rev.\  D {\bf 60} (1999) 014010;
  B.~Pire and L.~Szymanowski,
  %``QCD analysis of anti-p N --> gamma* pi in the scaling limit,''
  Phys.\ Lett.\  B {\bf 622} (2005) 83;
  %%CITATION = PHLTA,B622,83;%%
   B.~Pire, K.~Semenov-Tian-Shansky and L.~Szymanowski,
  %``A Spectral representation for baryon to meson transition distribution amplitudes,''
  Phys.\ Rev.\ D {\bf 82}, 094030 (2010) and
  %%CITATION = ARXIV:1008.0721;%%
  Phys.\ Rev.\ D {\bf 84}, 074014 (2011);
  %%CITATION = ARXIV:1106.1851;%%
  J.~P.~Lansberg, B.~Pire, K.~Semenov-Tian-Shansky and L.~Szymanowski,
  %``A consistent model for \pi N transition distribution amplitudes and backward pion electroproduction,''
  Phys.\ Rev.\ D {\bf 85}, 054021 (2012).
  %%CITATION = ARXIV:1112.3570;%%

  
  \bibitem{talkKirill} K. Semenov Tian-Shansky, talk at this conference,  arXiv:1206.6714 [hep-ph].
  %%CITATION = ARXIV:1206.6714;%%
  
  \bibitem{talkAnselmino} M. Anselmino, talk at this conference.
  
  \bibitem{talkGamberg}  L. Gamberg, talk at this conference, arXiv:1207.2444 [hep-ph].
  %%CITATION = ARXIV:1207.2444;%%
  
  
  
\end{thebibliography}
\end{document}